\documentclass[]{castellate_report}
\usepackage{xurl}
\usepackage{graphicx}
\usepackage{rotating}
\graphicspath{ {./plots/} }
\usepackage{blindtext}
\usepackage[skip=2pt,font=normalsize]{subcaption}
\usepackage[most]{tcolorbox} 
\usepackage{booktabs}
\usepackage[textcolor=blue, backgroundcolor=white, textsize=tiny, disable]{todonotes}
\usepackage{algorithm}
\usepackage{algpseudocode}
\newcommand{\VTNote}[1]{\todo{VT: #1}}
\newcommand{\ACNote}[1]{\todo{AC: #1}}

\newcommand{\commentOut}[1]{}

\definecolor{block-gray}{gray}{0.95}
\definecolor{block-lightblue}{RGB}{240, 248, 255}
\definecolor{block-darkblue}{RGB}{0, 0, 139}
\usepackage{footnote}
\BeforeBeginEnvironment{fquote}{}

\newtcolorbox{fquote}[1][]{%
    colback=block-lightblue,
    grow to right by=-10mm,
    grow to left by=-10mm, 
    boxrule=0pt,
    boxsep=0pt,
    breakable,
    enhanced jigsaw,
    borderline west={4pt}{0pt}{block-darkblue},
    #1,
}

\newtcolorbox[auto counter]{rec}[2][]{%
    colback=block-gray,
    grow to right by=-10mm,
    grow to left by=-10mm, 
    boxrule=0pt,
    boxsep=0pt,
    breakable,
    enhanced jigsaw,
    borderline west={4pt}{0pt}{logobg},
    title={Action~\thetcbcounter: #2\par},
    colbacktitle={block-gray},
    coltitle={black},
    fonttitle={\large\bfseries},
    attach title to upper={},
    #1,
}
\usepackage{hyperref}
\addto\extrasenglish{%
}
\addto\extrasenglish{%
}
\addto\extrasenglish{%
}
\usepackage{graphicx} 
\usepackage{url}
\usepackage{appendix}
\title{Security analysis of the Australian Capital Territory's eVACS 2020/2024 paperless direct recording electronic voting system}
\author{Chris Culnane\\chris@castellate.com \and Andrew Conway\\andrew@andrewconway.org \and Vanessa Teague\\vanessa.teague@anu.edu.au \and Ty Wilson-Brown\\twb@riseup.net}

\date{September 23, 2024}

\begin{document}

\maketitle

\commentOut{
Electoral Act 1992.
Division 9.3 Electronic voting and vote counting

118A Approval of computer program for electronic voting and vote counting

 (1) The commissioner may approve 1 or more computer programs for any of the following:

 (a) to allow electronic voting in an election;

 (b) to perform steps in the scrutiny of votes in an election.

 (2) The commissioner may approve a program under subsection (1) (a) only if the program will—

 (a) allow an elector to show consecutive preferences starting at ‘1’; and

 (b) give an elector an opportunity to correct any mistakes before processing the elector’s vote; and

 (c) allow an elector to make an informal vote showing no preference for any candidate; and

 (d) not allow a person to find out how a particular elector cast their vote.
}


Paperless Direct Recording Electronic (DRE) Voting machines are uncontroversial in most democracies because they are forbidden.  Independent studies across at least 4 continents have shown serious problems with both the integrity and privacy of the votes~\cite{aranha2019return, ca-top-to-bottom, gongrijp2007studying, wolchok2010security}, but it is the inherently unverifiable design---rather than specific vulnerabilities--- that has lead many legislators and electoral authorities to conclude that they should not be used.  One holdout is the Australian Captial Territory, which maintains (and has recently reimplemented) its paperless DRE system, known as eVACS, now used by $3/4$ of voters.

This report describes the implications for eVACS of two cryptographic errors in the Ada Web Services Library that it depends on. We identified these errors in the course of examining and testing the 2024 eVACS code, which was made publicly available in March 2024. We disclosed the problems to AdaCore, and explained the implications at the time to the relevant electoral authorities.

\autoref{sec:privacy} describes the implications of \href{https://cve.mitre.org/cgi-bin/cvename.cgi?name=CVE-2024-41708}{CVE-2024-41708}, in which a non-cryptographic pseudorandom number generator (PRNG) was used for shuffling votes. We demonstrate reversing the shuffle and hence recovering the order in which the votes were cast, which allows for identifying individual voters' choices if information on voting order is known to the adversary.
\autoref{sec:columnBias} describes a different implication of the way the PRNG was used: the randomised column selection for candidates is also slightly biased.

\autoref{sec:tls} examines the security of the TLS implementation, intended to maintain the integrity and confidentiality of votes as they are transmitted from the voting client to the server on the local network in a polling place. We found empirically that no TLS certificate validation was occurring in eVACS, implying that the connection was no more secure than a plaintext http connection. Again the issue
 was caused by underlying issues in the Ada Web Services library. The general issue is described in \href{https://docs.adacore.com/corp/security-advisories/SEC.AWS-0031-v2.pdf}{SEC.AWS-0031}; this report concentrates on the implications for vote integrity in eVACS.
 \ACNote{It is hard to tell but likely that that even if the Ada Webservices library supported name verification, they still didn't have a valid signing authority or possibly even name. So it was not necessarily caused by the Ada library. But it is
 hard to be sure. Certainly it was never tested.}

Ada Web Services have released updated library functions and documentation, which Elections ACT have incorporated into an updated version of the eVACS software, released in September 2024.\footnote{\url{https://www.elections.act.gov.au/__data/assets/file/0009/2571975/eVACS-Source-Code-15-Sept-2024.zip}} This is expected to run in the October 2024 election. 


Nothing about this report is intended to suggest that the system would be adequate for public elections if these specific vulnerabilities are patched. eVACS suffers from fundamental design flaws, particularly the absence of any opportunity for voters and scrutineers to verify the results, that remain unresolved. Although it is a step forward that the code has been made available in advance of the election, there is no way to verify that that code is properly installed on the voting devices or the voting server, nor that it functions properly on election day. The design does not produce evidence that the votes that determine the result are the accurately-recorded intentions of voters.

\section{Linking voting order to votes in eVACS 2020/2024} \label{sec:privacy}

\vspace{-10mm}
\textcolor{blue}{
\paragraph{Main finding (privacy)}
The \emph{pindex} value attached to each published vote can be used to recover the voting order. An attacker who learns individual voting order can therefore learn some individuals' votes.
}
\vspace{5mm}

The ACT \emph{Electoral Act 1992} requires e-voting systems to protect the secret ballot. Section 118A(2) states 
``The commissioner may approve a program under subsection (1) (a) only if the program will—
$\ldots$
 (d) not allow a person to find out how a particular elector cast their vote.''\footnote{\url{https://www.legislation.act.gov.au/View/a/1992-71/current/html/1992-71.html}}
 
In 2018 Ty Wilson-Brown identified a risk to the secret ballot for people who voted using the ACT's eVACS e-voting system: because electronic votes were stored sequentially, information about voting \emph{order} could be used to identify the individual's vote in the database.\footnote{\url{https://github.com/teor2345/Elections2018/blob/master/ElectionsACTDisclosure.md}} The publicly-released vote data also maintained this order, listing each batch of votes in the order they were cast. Wilson-Brown listed several example attacks, such as
\begin{itemize}
    \item linking the vote database with electronic roll mark-off data (possible for insiders or an attacker who compromises the system);
    \item finding one's own vote and identifying the people who voted immediately before or after (possible for a voter with a long preference list);
    \item noting the order in which people vote, particularly the first or last groups of voters (possible for anyone who observes the polling place).
\end{itemize}

Although Elections ACT and their software provider did not publicly acknowledge the problem,\VTNote{Is this too snarky?} they nevertheless attempted to ameliorate it. Public documents about the 2020 election stated that ``the vote order within the votes database will
be shuffled as part of the daily export process.''\footnote{\url{https://www.elections.act.gov.au/__data/assets/pdf_file/0009/1659798/Security-and-eVACS-v1.1.pdf}} However, the source code for the 2020 version of eVACS was not openly available, so we could not independently assess the security of the new solution at the time.

In March 2024, Elections ACT published the code that would be used for the October 2024 ACT election, which seems likely to be very similar to the code used in 2020.\VTNote{And then incompetently tried to un-publish it and then re-published it, and we've still got a long way to go.} It was immediately clear that the  shuffling algorithm is not secure---voting order can be easily recovered, again allowing voters to be linked to their votes in the database or the published votes, using the attacks described in 2018. We have demonstrated successful recovery of voting order on almost all votes cast via eVACS in 2020. 

\vspace{2mm}
For example, the third person to vote electronically on election day 2020 in Flynn gave their third preference to [*Redacted for public version*].

\paragraph{Recommendations}
First, it is important to identify potential responses we do \emph{not} recommend.
\begin{itemize}
    \item Do \emph{not} stop publishing the votes. Publishing the votes allowed serious errors in the count to be identified in 2020.\footnote{\url{https://github.com/AndrewConway/ConcreteSTV/blob/main/reports/2020\%20Errors\%20In\%20ACT\%20Counting.pdf}}
    \item Do \emph{not} stop publishing the source code. Publishing the code allowed for public examination and notification of a problem that was already there---it did not solve the problem to hide the code behind a confidentiality agreement in 2020, and it would not solve the problem to remove the code from public view again. It would simply mean that other similar problems would go unnoticed and hence uncorrected.
\end{itemize}

The main recommendation for addressing this privacy vulnerability is to reiterate the one made by Wilson-Brown in 2018:
\begin{description}
    \item[R1:] Randomise the order of all individual ballot paper preference data from electronic votes in future and previous elections.
\end{description}

The randomisation needs to be done using a cryptographic pseudorandom number generator. This generator must have a large, unpredictable seed, and generate an order that is not easily reversed. See \autoref{subsec:PRNGs}.
The gnat Ada library random number generator (currently used) is not suitable for this purpose, for three reasons:

\begin{description}

    \item[small seed] the random number generator is seeded using a 32 bit value, where approximately 3 of the bits are guaranteed to be zero. Best practice is to use 128 or 256 random bits, with no guaranteed ones or zeroes.
    \item[predictable sequence] the Ada ``Mersenne Twist'' random number generator produces a sequence that can be fully predicted after seeing around 600 outputs. So regardless of the seed, the vote order could still be found by collecting enough vote IDs. Instead, use a cryptographically secure alternative implementation.
    \item[predictable seed] the random number generator is seeded from the clock time when the system is started up.  In the current system, knowing the time the system was started is not particularly useful as the seed is so small and the time resolution so high (nanoseconds) that it overflows and wraps around approximately every 4.3 seconds. However, if a larger seed was used, which is necessary in any case, it could still be attacked if it could be predicted from the time.
\end{description}

\paragraph{Related work}
Insecure shuffling of votes has been observed in other electronic election systems, including Brazilian electronic voting machines~\cite{aranha2019return} \VTNote{That may not quite be the right reference} and some US vote-scanning machines~\footnote{\url{https://www.cisa.gov/news-events/ics-advisories/icsa-22-154-01}, \url{https://nvd.nist.gov/vuln/detail/CVE-2022-48506}, \url{https://dvsorder.org/}}.

\subsection{How eVACS generates the published votes} \label{sec:generatepindex}
\begin{figure}
    \centering
    \includegraphics[scale=0.2]{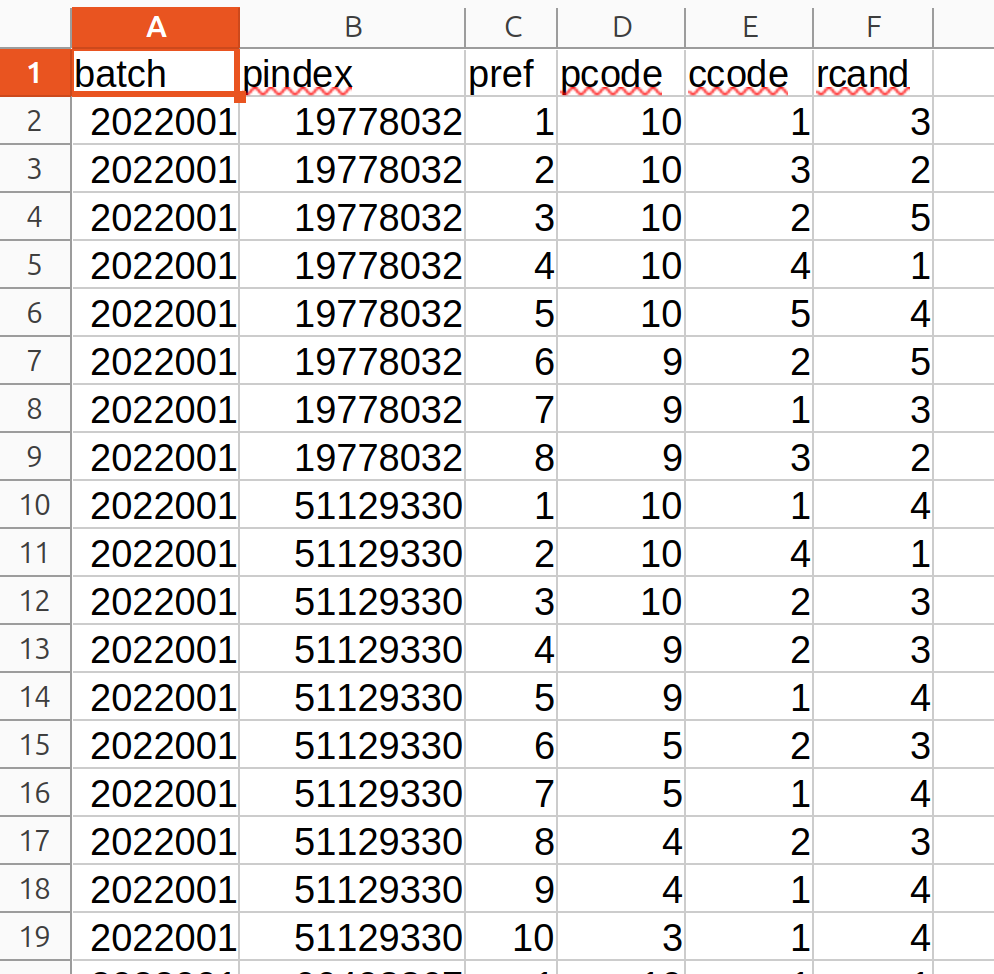}
    \caption{Two published votes from Ginninderra 2020. Note that each vote has several different preferences, all with the same \textsf{pindex}}
    \label{fig:Ginninderra}
\end{figure}

Elections ACT publishes detailed vote data for each election, including not only the individual preferences but also significant metadata about each ballot such as the location where it was cast, whether it was a paper or electronic vote, and a random-looking number called \textsf{pindex}. Figure~\ref{fig:Ginninderra} is an example, showing two votes cast in Gininderra in 2020.

Within each batch, votes are sorted by pindex. A document prepared by Software Improvements and published on the Elections ACT website\footnote{\url{https://www.elections.act.gov.au/__data/assets/pdf_file/0009/1659798/Security-and-eVACS-v1.1.pdf}} states that it is a requirement to ``Ensure that an elector and
their preferences cannot be
matched through the use of
timestamp data.'' This is addressed by ``shuffling the vote,'' though the exact mechanism is not described. It seems to be achieved by generating the pindex value, then sorting according to that value. This would be reasonable enough if the pindex values were truly random, but unfortunately the underlying random number generator in Ada is not adequate.

\subsubsection{Background: pseudorandom number generation} \label{subsec:PRNGs}
Although modern computers do have true sources of randomness, these are seldom used directly for generating random data values. Instead, ``random-looking'' values  are generated in a two-step process:
\begin{enumerate}
    \item first, a (hopefully) genuinely random value called the \emph{seed} is generated,
    \item subsequently, a series of values is generated in a deterministic process as a function of the seed.
\end{enumerate}
The values generated in this way are called \emph{pseudorandom}, because they ``look random'' but are actually a deterministic function of the initial seed. There are different technical definitions of what it means to ``look random''---cryptographic pseudorandom number generators guarantee that predicting the next values in the sequence, or recovering the seed, is computationally infeasible. However, non-cryptographic pseudorandom number generators achieve a weaker notion of ``random-looking'' which does not guarantee that the seed is hard to recover.  Once the seed is known, the complete sequence of values can be regenerated.

\subsubsection{How eVACS generates pseudorandom pindex values}

The eVACS Voting\_server.Random module uses Ada's \textsf{Ada.Numerics.Discrete\_Random} package to generate a value in the range from 0 to 999,999.


eVACS uses its implicit initialisation (Reset) option, which with the gnat library initialises the seed to a time-based value. The series of pseudorandom values are then generated via successive calls to get the next value.

The initialisation value has only 32 bits. Even if the initialisation time cannot be accurately guessed, the set of all $2^{32}$ possible seeds can be exhaustively enumerated.

Ada.Numerics.Discrete\_Random also offers an explicit initialisation option Reset(seed), which takes a 32-bit signed value. This produces the same set of initial states as the implicit initialisation option.\footnote{\url{https://docs.adacore.com/gnat\_rm-docs/html/gnat\_rm/gnat\_rm/implementation\_defined\_characteristics.html}}

\subsection{Algorithm for un-shuffling ballots}
The simple generation structure suggests a simple un-shuffling algorithm, which could be applied to each batch in turn. Try each possible seed and, for each seed, generate as many random values as there are ballots in the batch. Check whether the \emph{set} of generated values (in any order) matches the set of published pindex values for the batch.
If the sequence of generated values perfectly matches a (reasonably long) set of published pindex values, we can be confident that seed was used to generate those values, and that the order of voting matches the generation order in the algorithm.\footnote{False-positive matches are possible but highly improbable, especially for long sequences.}

When we tried this, we found that the generated values occasionally miss. We speculate that these correspond to test votes, or perhaps to voters who commenced a voting session but did not complete it, or to previously encountered pindex values produced a second
time and declared ineligible. We therefore allow a threshold MISS\_THRESHOLD of size comparable to the number of ballots (in practice
it is significantly smaller). 
eVACS votes are identifiable in the published vote data by their 000-terminated batch numbers---recovery obviously does not work for paper ballot batches.

Algorithm~\ref{alg:simple} describes an idealised version of the final algorithm.

\begin{algorithm}[H]
\caption{Seed recovery algorithm}\label{alg:simple}
\begin{algorithmic}
\For{$seed = 0$ to $2^{32}-1$} 
    \State Random\_Id.Reset(seed) 
    \State Unseen $\gets$ \{ set of known pindex values\} 
    \State Misses $\gets$ 0
    \While{Unseen is not empty and Misses<MISS\_THRESHOLD}
        \State $r\gets$ Random\_Id.Random
        \If{$r\in$Unseen}
            \State Unseen $\gets$ Unseen-\{$r$\}.
        \Else
            \State Misses $\gets$ Misses+1
        \EndIf
    \EndWhile
    \If{Unseen is empty}
        \State return seed. 
    \EndIf
\EndFor
\end{algorithmic}
\end{algorithm}

We actually use a slightly more complex version of this idealised algorithm:
\begin{itemize}
    \item For speed, only seeds divisible by 8 need checking, due to the way gnat library seeding works.
    \item For speed, one can check on a smaller number of ballots and give up on a seed when the number of misses
    gets too large (just trying the most successful seeds after ten ballots works).
    \item Some files had combined groups due to using a new seed on each day for early voting or bunching together
    multiple polling places. This required only looking for a run hitting a large number, but not all, of the known pindex values.
\end{itemize}

The resulting algorithm can test all seeds in a few minues on a home computer.

Once one has the seed, one can replay the order in which the pindex values were found in the above
algorithm to order the votes timewise.

\subsection{Results} \label{sec:results}
We were able to recover voting order for all batches of votes cast via eVACS in the 2020 ACT election, except the 327 in the Central Scrutiny, which seem to include a mix of smaller collections of votes cast elsewhere. 

For the eVACS votes cast on polling day, for each location (other than Central Scrutiny), a single seed was sufficient to generate all the pindex values. The number of skipped pindex values tended to be very small.

For eVACS votes cast through the early voting period (19 days), every location required 19 seeds to generate all pindex values, except Dickson and Kippax, which both required 20. This is consistent with the server being restarted daily, and also an extra time in each of those two locations.

\section{Bias in the generation of random column selections} \label{sec:columnBias}

\vspace{-10mm}
\textcolor{blue}{
\paragraph{Main finding (party bias)}
The initially selected column is biased by up to 1.6\% towards some political parties. This impacts audio voting at polling booths and over the phone.
}
\vspace{5mm}

As part of electronic voting, the initially selected column on the ballot is randomised. Unfortunately, the implementation is biased towards the leftmost parties on the ballot. This issue is separate from the privacy issues described in the rest of this report.

Each column represents a political party, or ungrouped independent candidates. The order of the party columns on all ballot papers is chosen randomly, as part of the ballot draw. The order of candidates within a column is also rotated between different ballot papers, using the Robson Rotation. These processes are used for both paper and electronic voting. We are not aware of any bias in these processes. 

However, electronic voting also randomises the initially selected column of candidates, to further reduce bias. This randomisation is used during audio voting with physical keypads and over the phone. It is unclear whether it also happens for touch screen voting.

The implementation takes a large pseudo random number, gratuitously\footnote{Nothing is gained by the explicit effort of making it small. The only effect is to increase the size of the bias.} 
converts it to a small random number $r$ between 0 and 255 and then takes
that number modulo the number of parties $p$ on the ballot. 
That is, divide $r$ by $p$ and only consider the remainder. Suppose there
were 10 parties ($p=10$). Then if $r$ were 0, 10, 20,..., 250, then the
first party would be chosen. If $r$ were 1, 11, 21,..., 251, then the second
party would be chosen. However $r$ will never be 256, so the seventh and subsequent parties
will each get one fewer opportunity to be chosen than the first six parties. 

The exact impact depends on the number of columns. For 2020, the resulting biases were\footnote{Source: https://www.elections.act.gov.au/elections\_and\_voting/2020\_legislative\_assembly\_election/list-of-candidates}:
\begin{description}
    \item \textit{8 columns on the ballot, divides evenly into 256}
    \item[Brindabella] no bias
    \item \textit{11 columns on the ballot}
    \item[Ginninderra] bias of 1.2\% (3/256) towards 3 unelected parties
    \item \textit{all remaining electorates have 9 columns on the ballot}
    \item[Kurrajong] bias of 1.6\% (4/256) towards Greens, Liberals, and 2 unelected parties
    \item[Murrumbidgee] bias of 1.6\% towards Greens and 3 unelected parties
    \item[Yerrabi] bias of 1.6\% towards Greens, Liberals, and 2 unelected parties
\end{description}

This is not a large source of bias, and will not impact a large portion of voters. However it is entirely
unnecessary, and the margins in ACT elections are very small, so it is difficult to tell what impact this had.

\ACNote{I have removed the following as I feel like it will be easy for the Elections ACT to confuse the
issue by saying that it only affected a smaller number of people (say only audio) and therefore our
analysis is all wrong and it is insignificant. However we can uncontestably say that it is a gratuitous
bias. Removed: 
In each of the 4 biased electorates, the total bias is larger than the winning margin for one or more candidates. However, the ballot column order also introduces a bias, and some voters won't be impacted by the initial column selected by the system. So it is difficult to tell how much this bias impacted the outcome of the 2020 election.
}

It would have been easy to produce unbiased initial columns as follows: 
\begin{description}
    \item[Use tested library APIs] Configure Ada's pseudo random number generator using the exact range of columns on the ballot paper for each electorate. Its implementation is specifically designed to produce unbiased results. Any further calculations on the random values can introduce bias.
\end{description}

\section{TLS Security} \label{sec:tls}

\vspace{-10mm}
\textcolor{blue}{
\paragraph{Main finding (network security)}
Polling place voting booth clients used a software library which did not check it was talking to the correct server. This meant network intruders could easily collect or modify votes.
}
\vspace{5mm}

EVACS conveys vote selections from voting clients to a server via the polling station's local area network (LAN). Earlier versions of eVACS used an unencrypted http connection, trusting that the network was perfectly secure, but from 2020 the developers added the TLS protocol to protect these transmissions. TLS is the same protocol that protects most ordinary web-browsing on the modern Internet. It has three main functions.
\begin{enumerate}
\item {\bf Server authentication: } the client can verify the identity of the server.
\item {\bf Data confidentiality: } the contents of the communication cannot be read by other parties on the network.
\item {\bf Data integrity: } the contents of the communication cannot be modified by other parties on the network.
\end{enumerate}
All these properties are crucial for voting: votes must be sent to the correct server, must not be visible to others, and must not be alterable. Importantly, the first property is crucial for achieving the other two: if the client does not carefully check that it is sending the information to the \emph{correct} server, then any other integrity or privacy guarantees are irrelevant.

\emph{TLS certificate validation} is the process in which a client verifies a server's identity by checking its TLS certificate. The details are complicated, but the main things to check are:
\begin{enumerate}
    \item {\bf validity: } that the certificate is valid (e.g. has not expired and has a valid digital signature),
    \item {\bf trust: } that the certificate is signed by a trusted Certificate Authority (CA),
    \item {\bf hostname validation: } that the certificate belongs to the expected host. 
\end{enumerate}

When we examined the publicly-available version of eVACS we noted that the TLS certificate was not correctly formed to allow hostname validation, specifically, it did not contain a hostname in any of the applicable attributes. This provided a strong indicator that the certificate validation was either not functioning correctly or was incorrectly configured.

Further examination of the publicly-available source code of eVACS showed that eVACS was using the default configuration for the Ada Web Server (AWS) Client component. Switching our analysis to the AWS Client, we created a series of test cases that sought to test the TLS certificate validation. We discovered that the default configuration did not correctly handle certificate validation. More specifically, validation errors returned from OpenSSL were silently ignored. This included expired certificates, untrusted certificates, and hostname validation errors\footnote{\url{https://badssl.com} contains a variety of bad TLS certificates for testing.}. 

Whilst there was an esoteric way of configuring the AWS Client to capture some validation errors, it required including a superfluous client TLS certificate to trigger the necessary certificate checking configuration option. Furthermore, there was no way of triggering hostname validation as the necessary functionality was not implemented within AWS.

As such, eVACS was operating without any form of TLS certificate validation. This implies that any attacker present on the network, if they could intercept the unencrypted TLS handshake, could spoof the server and collect, read and alter votes. 

Our discovery was responsibly disclosed to AdaCore, since this issue impacted any users of TLS in the AWS Client. AdaCore confirmed the existence of the vulnerability, as described in \href{https://cve.mitre.org/cgi-bin/cvename.cgi?name=CVE-2024-37015}{CVE-2024-37015}, and implemented and released a fix\footnote{AdaCore security updates are posted to \url{https://www.adacore.com/cybersecuritycenter}.}.

Whilst the underlying issue was in the Ada Web Server Client component, this does not explain why the lack of certificate validation was not picked up during testing or auditing. It was obvious to us that the certificate validation was not functioning correctly, or at least conventionally, because the contents of the TLS certificate lacked the necessary attributes. This should have been a trivial find for the auditors. Furthermore, test cases within the code should have included testing what happened when an invalid certificate was presented by a server.


\section{Code updates 15 Sep 2024}
Following the disclosure of these results to Elections ACT and AdaCore, Elections ACT released an updated version of the code, which incorporated AdaCore's updated libraries for both PRNGs and TLS. We did not test this thoroughly---indeed, thorough testing is not possible because the published code does not include everything necessary to get the code running. 

Additionally, the 15th September 2024 code release no longer contains sample certificates. This prevents examination of their construction to check appropriate attributes have been set. 

We also note that the AWS Client implementation sets a client TLS certificate, despite the server not being configured to use Mutual TLS. As such, the client certificate is never used. Whilst this was a necessary workaround to trigger certificate validation before the AWS update, it is not necessary with the updated code. There is no explanation of how the client certificate is generated, or whether there are individual certificates on each client. This does not constitute a security vulnerability, it is just a redundant certificate, however, it raises questions about the level of understanding of the code and the configuration options.

\section{Conclusion: Implications and recommendations}

Openly releasing the source code early in 2024 gave us time to report these vulnerabilities before the 2024 election.

Unfortunately, in 2020 there was a strict NDA, which prevented us from doing a similar analysis. This delayed the discovery of counting bugs until the open release of the full preference data, and the column bias and vote ID privacy problems until the 2024 code was openly released.

The bias in the selection of random columns (\autoref{sec:columnBias}) can be easily corrected, and changes have been made in the code released on 15 September. 

Unfortunately, there is no straightforward fix to the privacy problem (Sections~\ref{sec:generatepindex} to~\ref{sec:results}). Although secure shuffling (which seems to be implemented now, though we have not checked thoroughly) may protect \emph{future} elections, the votes exposed in the past cannot be un-exposed. It may also still be possible in future to derive timing information from the raw database log files, which may still be accessible to insiders.
This is important as a large portion of the
point of a secret ballot is keeping people's votes secret from the authorities. Insiders are also
the people who have the easiest time linking individuals to vote order.


The system's design is also fundamentally insecure. Even ignoring the TLS issues 
described above, the system suffers from a fundamental design flaw: it is not designed to produce evidence that the published votes are the accurately recorded and properly processed intentions of the voters.
Undetectable failures could
be due to bugs, hardware errors, operator errors, deliberate manipulation, insiders or supply chain attacks.

A voting system needs to be secure and private. The eVACS voting system is neither, and has no 
reasonable prospect of being made so. It is not fit for purpose, and does not apply democratic best practices. Prior runs---up to and including 2020--- demonstrably did not meet legislative requirements.

Using eVACS in an election means that if there is a disputed election result, it will be difficult to prove that the system correctly collected and counted votes, in an unbiased way.

While there will always be some risk in any election system, the risks of eVACS are significantly higher than a paper ballot. It should not be
used in an election. 
\bibliographystyle{alpha}
\bibliography{bib}

\end{document}